\documentclass[prl,12pt,onecolumn,superscriptaddress]{revtex4-1}
\usepackage{amsmath,amssymb,graphicx,color,xfrac}

\usepackage[mathletters]{ucs}
\usepackage[utf8x]{inputenc}
\usepackage{ulem}

\definecolor{lightblue}{rgb}{0.17,0.39,1}
\definecolor{lightgreen}{rgb}{0.67,0.81,0.08}
\definecolor{lightred}{rgb}{1,0.05,0.52}

\usepackage[
bookmarks,
colorlinks=true,
breaklinks,
pdftitle={Robust spin correlations at high magnetic fields in the honeycomb iridates},
pdfauthor={Kim,Brad,Ross,Arkady}
]{hyperref}
\hypersetup{linkcolor=lightblue,citecolor=lightblue,filecolor=black,urlcolor=lightblue}

\newcommand{\sign}{\ensuremath{\text{sign}}}

\begin{document}
\title{Robust spin correlations at high magnetic fields in the honeycomb iridates}

\author{K. A. Modic}
\email[Email: ]{modic@cpfs.mpg.de}
\affiliation{Max-Planck-Institute for Chemical Physics of Solids, Noethnitzer Strasse 40, D-01187, Dresden, Germany}
\affiliation{Los Alamos National Laboratory, Los Alamos, NM 87545, USA}
\author{B. J. Ramshaw}
\affiliation{Los Alamos National Laboratory, Los Alamos, NM 87545, USA}
\affiliation{Laboratory for Atomic and Solid State Physics, Cornell University, Ithaca, NY 14853, USA}
\author{Nicholas P. Breznay}
\affiliation{Materials Science Division, Lawrence Berkeley National Laboratory, Berkeley, California 94720, USA}
\affiliation{Department of Physics, University of California, Berkeley, California 94720, USA}
\author{James G. Analytis}
\affiliation{Materials Science Division, Lawrence Berkeley National Laboratory, Berkeley, California 94720, USA}
\affiliation{Department of Physics, University of California, Berkeley, California 94720, USA}
\author{Ross D. McDonald}
\affiliation{Los Alamos National Laboratory, Los Alamos, NM 87545, USA}
\author{Arkady Shekhter}
\affiliation{National High Magnetic Field Laboratory, Florida State University, Tallahassee, FL 32310, USA}

\begin{abstract}
\end{abstract}

\pacs{...}
\date{\today }
\maketitle

\section{Abstract}

The complexity of the antiferromagnetic orders observed in the honeycomb iridates is a {\it double-edged sword} in the search for a quantum spin-liquid ground state: both attesting that the magnetic interactions provide many of the necessary ingredients, but simultaneously impeding access. As a result, focus has been drawn to the unusual magnetic orders and the hints they provide to the underlying spin correlations. However, the study of any particular broken symmetry state generally provides little clue as to the possibilities of other nearby ground states \cite{Anderson}.  Here we use extreme magnetic fields to reveal the extent of the spin correlations in $\gamma$-lithium iridate. We find that a magnetic field with a small component along the magnetic easy-axis melts long-range order, revealing a bistable, strongly correlated spin state.  Far from the usual destruction of antiferromagnetism via spin polarization, the correlated spin state possesses only a small fraction of the total moment, without evidence for long-range order up to the highest attainable magnetic fields ($>$90~T).

\section{Introduction}

Spin systems with highly anisotropic exchange interactions have recently been proposed to host quantum spin-liquid states \cite{Kitaev}. It has been suggested that the extreme exchange anisotropy required to achieve such states can be mediated by the strong spin-orbit interactions of transition-metal ions situated in undistorted and edge-sharing oxygen octahedra (Figure \ref{fig:4}d) \cite{JK}. Both the 2D ($\alpha$) and 3D ($\beta$ and $\gamma$) polymorphs of the honeycomb iridates (A$_2$IrO$_3$, A = Na, Li) closely fulfill these geometric requirements and display intriguing low-field magnetic properties that indirectly indicate the presence of anisotropic exchange interactions between the iridium spins \cite{Chun, Modic, Takagi}. However, the particular bond-specific Ising exchange interactions considered in the Kitaev model have not been directly verified. To date, all honeycomb iridates deviate sufficiently from this ideal, such that they transition to long-range magnetic order at finite temperature \cite{Radu, Radu2}, drawing attention to the complex magnetic structures of their ground states \cite{sodium, zigzag, zigzag2, order1, unified, spiral}. Just as the strange metallic state near the quantum critical point in the hole-doped cuprates is revealed once superconductivity is suppressed with magnetic field \cite{Brad}, we use high magnetic fields to destroy the antiferromagnetic order and expose a correlated ``spin-fluid" in $\gamma$-lithium iridate.

\begin{figure}[htbp]
\centering
\includegraphics[width=1.1\linewidth, trim=3cm 0cm 1cm 0cm, clip=true]{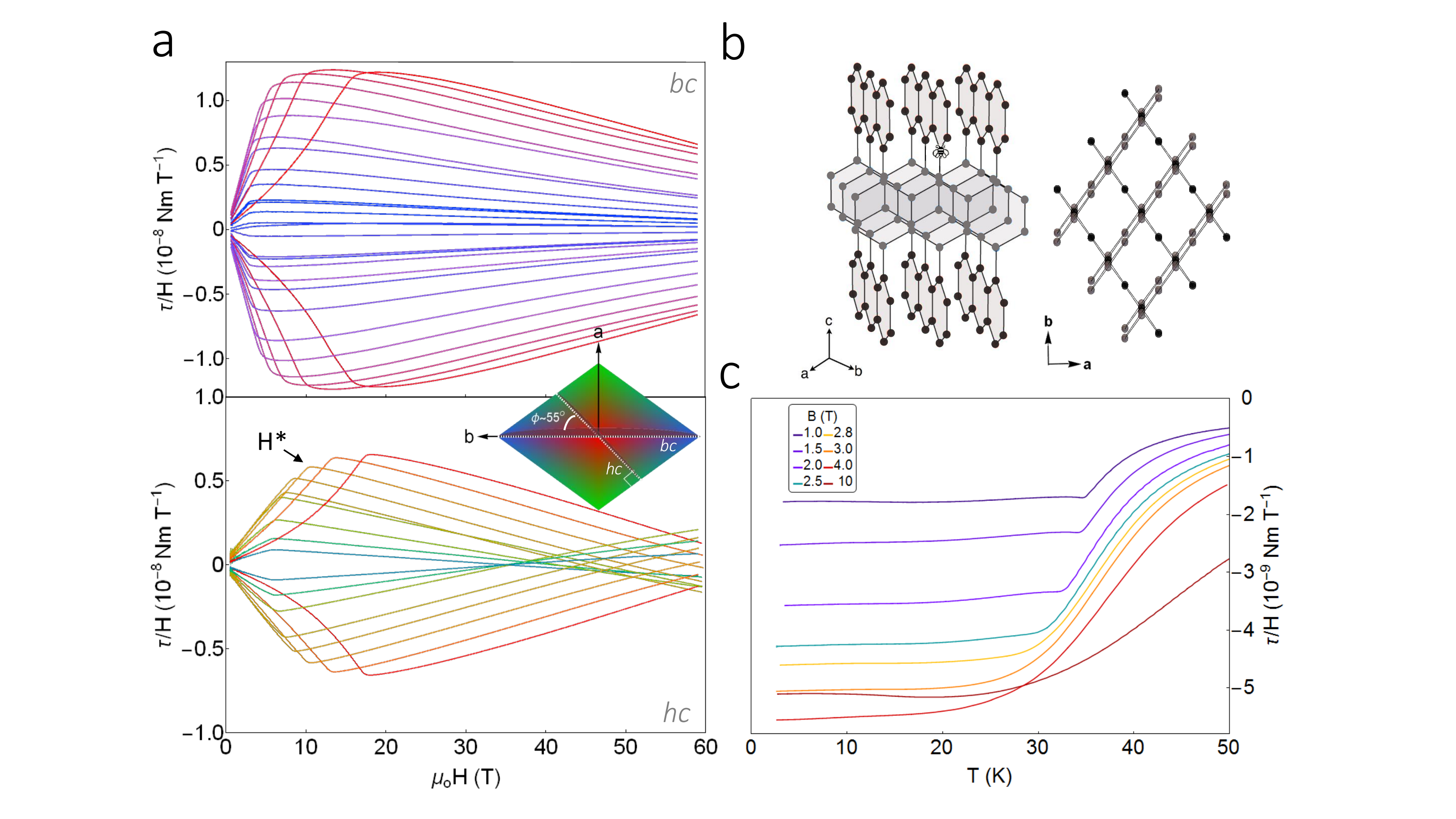}
\rule{35em}{0.5pt}
\caption[$τ/H$ at 100 Tesla]{(a) At 4K,  $\tau/H$ represents magnetic anisotropy as a function of magnetic field $H$ in the \textit {bc} (top)- and \textit{hc} (bottom)-planes.  Colored curves correspond to magnetic field applied along crystallographic directions depicted in the diamond-shaped schematic of the crystal morphology.  The crystal is rotated in the planes designated by white lines with respect to the applied magnetic field.  (b) 3D honeycomb structure of $\gamma$-lithium iridate.  View along the \textit{c}-axis illustrates the orientation of two honeycomb planes with respect to the crystal directions.  (c) $\tau/H$ as a function of temperature show a crossover from a sharp transition to long-range order at fields below $H^*$ to smooth change in torque that characterizes the onset of spin correlations at fields above $H^*$.}
\label{fig:1}
\end{figure}

$\gamma$-lithium iridate features a complex, yet highly-symmetric incommensurate magnetic structure with non-coplanar and “counter-rotating” moments below $T_N$ = 38 K \cite{Modic}. The magnetic anisotropy within the ordered phase was extensively characterized in our previous study \cite{Modic}.  With increasing magnetic field below $T_N$, the magnetic torque τ divided by the applied field H increases linearly up to an angle-dependent field $H^*$ (Figure \ref{fig:1}a), defining the boundary of the low-field, ordered state.  Importantly, this sharp feature is not accompanied by full saturation of the effective spin-1/2 iridium moment \cite{Modic}. Instead, the magnetic moment at $H^*$ is only 0.1 $\mu_B  $ \cite{Modic,Takagi} with fields applied along the magnetic easy-axis. The lack of a fully saturated moment at $H^*$ implies either the onset of another magnetically ordered phase above $H^*$, or alternatively, a transition into a paramagnetic state lacking long-range order, with spin correlations controlled by exchange interactions much stronger than the applied field.  

\section{Results}

We use torque magnetometry to examine the non-linear magnetic response of single crystal $\gamma$-lithium iridate at high magnetic fields.  The extreme anisotropy of $\gamma$-lithium iridate necessitated that both smaller volume samples and stiffer cantilevers were employed for the pulsed high-field measurements than for our prior DC low-field measurements \cite{Modic}. To this end, we utilized FIB lithography to cut and precisely align samples on Seiko PRC120 piezoresistive levers. This had the added benefit of reducing the lever deflection with field which can provide a systematic angle offset particularly close to the magnetically hard-axes.  

\begin{figure}[htbp]
\centering
\includegraphics[width=1.02\linewidth, trim=0.5cm 3.8cm 0cm 3.5cm, clip=true]{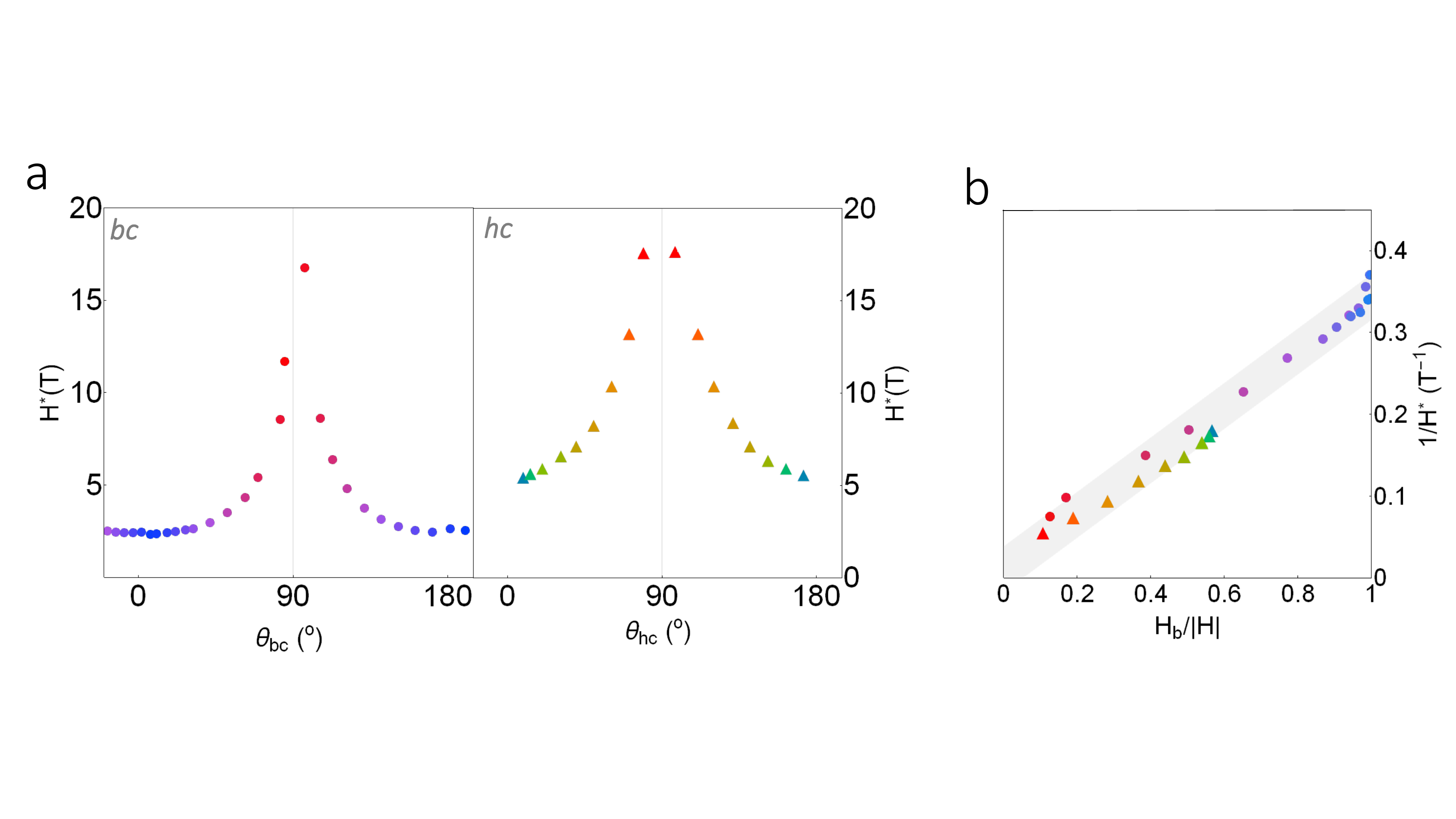}
\rule{35em}{0.5pt}
\caption[$τ/H$ for several field orientations in the \textit{bc}- and \textit{hc}-planes]{(a) $H^*$ versus rotation angle $\theta$, defined from the \textit{ab}-plane.  (b) When plotted against $H_b/|H|$, the collapse of 1/$H^*$ for field rotation in both planes illustrates that $H^*$ depends entirely on $H_b$.}
\label{fig:2}
\end{figure}

Figure \ref{fig:1}a shows $τ/H$ for field rotation in two planes that include the \textit{c}-axis: the \textit{bc}-plane ($φ=0^{\circ}$) and a plane that rotates in a direction perpendicular to one of the honeycomb planes, referred to as the \textit{hc}-plane ($φ=55^{\circ}$).  We find that $H^*$ closely follows $1/|\cos \theta\cosφ|$ (Figure \ref{fig:2}a), where $\theta$ denotes the angle between the \textit{ab}-plane and the applied field in both rotation sets. The collapse of $1/H^*$ versus the normalized $\textit {b}$-component of magnetic field $H_b/|H|$ onto a straight line in both rotation planes (Figure \ref{fig:2}b) indicates that the torque at $H^*$ is dominated by the $\textit {b}$-component of magnetization $M_b$ and that the value of magnetization $M(H^*)$ is nearly independent of field orientation. Moreover, the absolute value in the denominator of the angle dependence indicates a bistable behavior, $M_b =M_b^*\sign(H_b)$. To determine whether the bistable character of $M_b$ persists to fields above $H^*$, we turn to the angle dependence of $\tau/H$ across the entire field range.

\begin{figure}[htbp]
\centering
\includegraphics[width=1.0\linewidth, trim=6cm 2cm 6cm 2cm, clip=true]{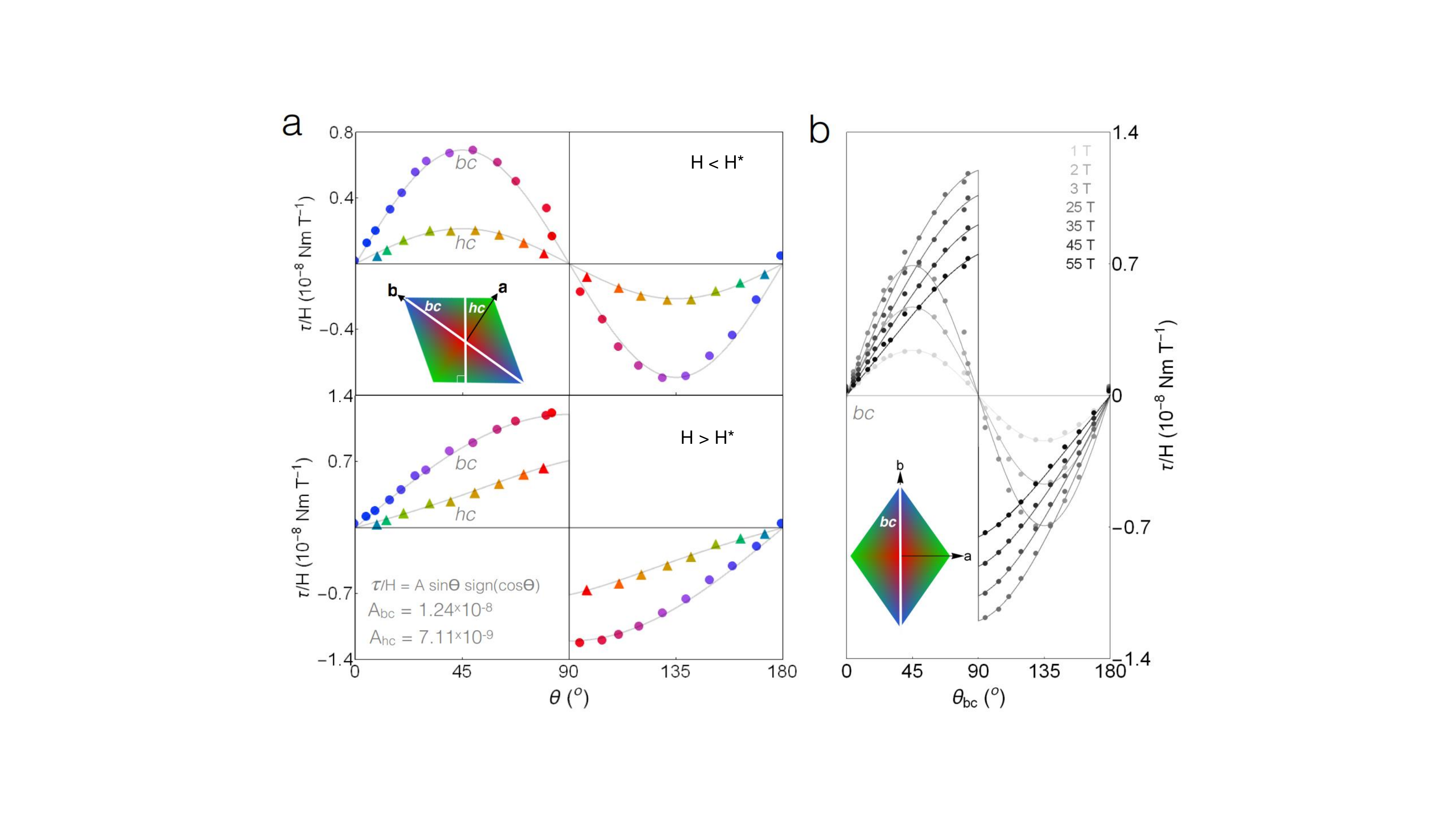}
\rule{35em}{0.5pt}
\caption[Field evolution of the angle dependence of $τ/H$ in the ${bc}$ and ${hc}$ planes]{Evolution of the angle dependence of $\tau/H$ in the ${bc}$- and ${hc}$- planes with increasing magnetic field.  (a) At 4K, $\tau/H$ evolves from a characteristic $\sin 2 \theta$ angle dependence at small fields to a correlation-driven $\sinθ$ $\sign(\cosθ)$ angle dependence at high fields.  (b) $\tau/H$ at discrete field values shows a deviation from $\sin 2 \theta$ behavior persisting to 55T.}
\label{fig:3}
\end{figure}

At fields below $H^*$, $τ/H$  follows a $\sin2\theta$ dependence that is characteristic of the linear response regime $ M_{i} = χ_{ij} H_{j}$ (Figure \ref{fig:3}a).  In this regime, torque vanishes for fields applied along the high symmetry directions and the amplitude of the $\sin2\theta$ angle dependence represents the anisotropy of the magnetic susceptibility $\alpha_{ij}=\chi_i-\chi_j$ \cite{Modic}.  Deviations from the smooth $\sin2\theta$ angle dependence provide a direct probe of magnetic correlations that are otherwise masked by the zero-field antiferromagnetic order.  
The angle dependence below $H^*$ is in stark contrast with the angle dependence at high fields where torque exhibits a sharp discontinuity as the magnetic field crosses the \textit{c}-axis (Figure \ref{fig:3}a). We observe the non-linear susceptibility at fields above $H^*$ as a $\sin\theta$ component with $\sign(\cos\theta)$ capturing the discontinuity at $90^\circ$. The $\sin\theta$ angle dependence indicates that the torque $τ=M_bH_c-M_cH_b$ in this high-field regime is dominated by the first term, where $H_c=H\sin \theta$.  The negligible contribution of the second term $M_cH_b$ is confirmed by the observed torque in the \textit{hc}-plane: if the discontinuity in $τ/H$ versus $\theta$ is driven exclusively by the saturated \textit{b}-component of magnetization, then one would expect the amplitude of $\sin \theta$ in the \textit{hc}-plane to be reduced by a factor $\cos55^{\circ}≈0.577$ compared to the \textit{bc}-plane.  Indeed, the reduced amplitude factor is 0.573, providing evidence that $M_B$ dominates the $\textit{total}$ magnetization in the high-field phase.  Futhermore, the discontinuity $M_b=M_b^*\sign(H_b)$ across the $\textit{c}$-axis implies that the Ising-like degeneracy extends well beyond $H^*$.

In the ultra high-field limit, when all spins are nearly aligned (and therefore all correlations are overcome), one expects to recover a simple $\sin2\theta$ angle dependence of the torque \footnote{ At very high fields all spins are nearly aligned with applied field. The effective local anisotropy energy (per formula unit) is $E ≈ (μ_\text{Ir}^2β/2)\cos2θ$ where $μ_{\text{Ir}}$ is magnetic moment of Ir-ion and $β$ is local anisotropic spin stiffness. The torque, $τ=-dE/dθ ≈ μ_\text{ir}^2β\sin2θ$ has $\sin2\theta$ angular dependence, similar to that at low fields.}. By contrast, in $\gamma$-lithium iridate, the $\sin\theta$ $\sign(\cos\theta)$ angle dependence of the torque persists up to the highest applied magnetic fields (Figure \ref{fig:3}b), indicating the presence of robust spin correlations throughout the entire field range and providing a lower bound for the magnitude of the exchange interactions.

\begin{figure}[htbp]
\centering
\includegraphics[width=1.05\linewidth, trim=0.5cm 0cm 0cm 0cm, clip=true]{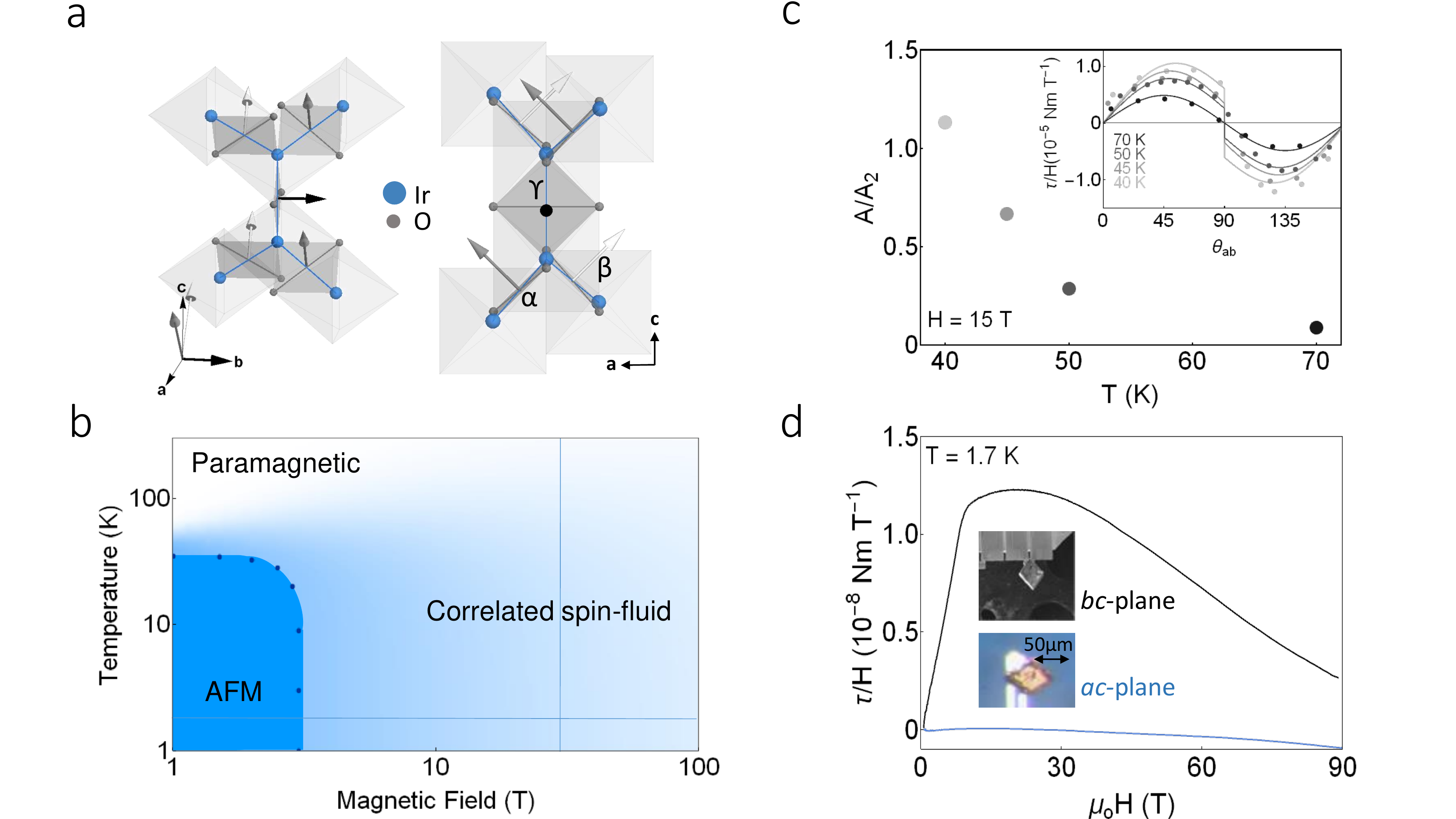}
\rule{35em}{0.5pt}
\caption[]{ (a)  Iridium-iridium exchange pathways mediated by oxygen octahedra.  "Kitaev" components of the exchange interactions are represented by orthogonal arrows between the nearest neighbor iridium-iridium bonds ($\alpha$, $\beta$ and $\gamma$).  (b) Schematic of the temperature/magnetic field phase diagram of $\gamma$-lithium iridate.  At low temperatures, the ordered phase (dark blue) extends to H = 3 T with a correlated spin state (light blue) likely persisting to magnetic fields above 100 T.  The faint lines represent regions of the phase diagram explored with constant field and temperature values in (c) and (d), respectively.  (c) Ratio of the coefficients in the fits of $\tau/H=A\sin\theta\sign(\cos \theta)+ A_2\sin 2 \theta$ at fixed temperatures (inset).  (d) $\tau/H$ shown for two principal components of magnetic anisotropy ($\alpha_{bc}$ and $\alpha_{ac}$) up to 90 T, highlighting the special role of $M_b$.}
\label{fig:4}
\end{figure}

To assess the extent of this correlated spin-fluid, we examine the thermal evolution of the non-linear angle dependence of torque at fields above $H^*$ (vertical line in Figure \ref{fig:4}b). Figure~\ref{fig:4}c shows the angle dependence of $\tau/H$ at 15~T for a range of temperatures, revealing a gradual decrease of the non-linear response as temperature is raised. The non-linearity becomes undetectable at this magnetic field above 70~K, the same temperature where paramagnetic behavior onsets in magnetic susceptibility measurements \cite{Modic}, indicating a continuous crossover to a strongly correlated spin-fluid at low temperatures. We note that specific heat measurements also observe a broad peak upon cooling at fields above $H^*$, consistent with the decrease in entropy due to the onset of spin correlations \cite{Alejandro}.  The onset of spin correlations as a precursor to long-range order is a common phenomena in low dimensional magnets (characterized by a large \textit{spatial} anisotropy of the exchange interactions), where the energy scale of the dominant exchange interactions is crossed at a temperature significantly above the transition \cite{yoshi, Blundell}.  We observe similar phenomena in $\gamma$-lithium iridate, but in this case, it results from the anisotropy in the \textit{spin}-components of the exchange interactions \cite{Modic}.

\section{Discussion}

The orientation of the spin-anisotropic exchange interactions with respect to the crystal directions
 gives rise to a strong magnetic anisotropy in the ordered state \cite{Modic}. The extreme softening of $\chi_b$ occurs because the  \textit {b}-axis is the only direction coaligned either parallel or perpendicular to all Ir-O$_2$-Ir planes.  None of the Ir-O$_2$-Ir planes are parallel or perpendicular to either the \textit{a}- or \textit {c}-axis (Figure~\ref{fig:4}a).

In this study, we find that the \textit{b}-component of magnetization continues to dominate the magnetic response beyond the ordered state to the highest measured fields (Figure \ref{fig:4}d). This unusual behavior is a direct indication that the anisotropy of the strongly correlated spin-fluid is driven by the exchange interactions, rather than \textit{g}-factor anisotropy tied to the honeycomb planes. This observation, coupled with the anomalously small magnetic moment induced for all field orientations, identifies that the spin correlations persist over a broad field and temperature range.  The high-field anisotropy resembles, but is not specific to, the zero-field broken symmetry state at low temperatures.  The commonality of the anisotropy in the ordered phase and the spin-fluid suggests that the correlated object is a sub-unit of the complex magnetic structure observed at zero field.  Perhaps the counter-rotating spin spirals \cite{Radu} are decoupled by a small component of field along the $\textit {b}$-direction, leading to a finite correlation length with only minimal net spin polarization.

A spin-liquid ground state is characterized by a lack of long-range order and gapless excitations \cite{Balents}. In this context, we note that many conventional, as well as correlated, metals are unstable at low temperatures and undergo symmetry breaking that gaps their low-energy excitations \cite{Matthias,Zaanen}. In frustrated magnets, long-range order can be stabilized by lattice distortions, crystal-field effects, and alternative exchange pathways. In the specific case of embedding the Kitaev model onto a 3D honeycomb lattice, it appears that long-range order is stabilized by intrinsic symmetry breaking – whereby only one component of the spin-anisotropic exchange can be coaligned with a crystal direction. An intriguing question for future studies is whether the high-field state of $\gamma$-lithium iridate inherits any properties of a gapless spin-liquid, such as the linear thermal transport in the zero-temperature limit observed in geometrically-frustrated spin systems \cite{thermal, spinon}.

\section{Acknowledgments.}

This work was performed at the National High Magnetic Field Laboratory, and was supported by the US Department of Energy BES “Science at 100 T”,  the National Science Foundation DMR-1157490, and the State of Florida. 

Correspondence and request for materials should be addressed to K.M. at modic@cpfs.mpg.de.

\bibliography{Refs-fixed}

\cleardoublepage

\end{document}